\documentclass[fleqn,twoside]{article}
\usepackage{espcrc2}

\usepackage{graphicx}

\newcommand{\bi}{\begin{itemize}}
\newcommand{\ei}{\end{itemize}}
\newcommand{\be}{\begin{eqnarray}}
\newcommand{\ee}{\end{eqnarray}}
\newcommand{\kp}{{\bf k}_\perp}
\newcommand{\pp}{{\bf p}_\perp}

\newcommand{\AmS}{{\protect\the\textfont2
  A\kern-.1667em\lower.5ex\hbox{M}\kern-.125emS}}

\title{$B$ Mesons on the
Transverse Lattice}

\author{M. Burkardt \address[NMSU]{
Department of Physics, New Mexico State University,
Las Cruces, NM 88011, U.S.A.}
\address[TU]{Physik Department T-39, TU M\"unchen,
D-85747 Garching, GERMANY}
        \thanks{
This work was supported by a grant from DOE (FG03-95ER40965) and through
Jefferson Lab by contract DE-AC05-84ER40150 under which the Southeastern
Universities Research Association (SURA) operates the Thomas Jefferson
National Accelerator Facility.} and
        Sudip K. Seal \addressmark[NMSU]}
       
\begin{document}

\begin{abstract}
We present results from a first study of 
$B$-mesons that is
based on a transverse lattice formulation of
light-front QCD. The shape of the Isgur-Wise form
factor is in very good agreement with experimental
data. However, the calculations yield rather large
values for $f_B$ and $\bar{\Lambda}$ compared to
contemporary calculations based on other techniques. 
\vspace{1pc}
\end{abstract}

\maketitle

\section{Introduction}
Parton distributions measured in deep inelastic scattering, as well 
as many other high-energy observables
(e.g. the deeply virtual Compton scattering amplitudes), are dominated 
by correlations along  
the light-cone ($x^2=0$). This simple fact poses a big obstacle
for non-perturbative calculations of these important observables.
For example, this makes direct evaluations of parton distributions on
a Euclidean lattice, where all distances are
space-like, impossible and calculations performed in a Euclidean
framework usually try to reconstruct parton distribution functions 
from their moments.
Furthermore, in an equal time quantization scheme,
deep inelastic structure functions are described by
real time response functions which are not only
very difficult to interpret but also to calculate.

Light-Front (LF) quantization seems is a promising
tool to describe the immense wealth of experimental
information about structure functions for a variety
of reasons:
\begin{itemize}
\item correlations along the light-cone become ``static'' observables
in this approach
[i.e. equal $x^+ \equiv (x^0+x^3)/\sqrt{2}$ observables] 
\item structure functions are easy to evaluate from the 
LF wavefunctions
\item structure functions are easily interpreted as LF 
momentum densities
\end{itemize}
Further advantages of the LF formalism derive from the
simplified vacuum structure (nontrivial vacuum effects
can only appear in zero-mode degrees of freedom) which
provides a physical basis for the description of
hadrons that stays close to intuition
\cite{nato,bigguy,osu,mb:adv}.\\[2.ex]
\section{The Transverse Lattice}
Before one can apply the LF formalism to QCD one has to remove
the divergences first (i.e. regularize and renormalize).
Then one has to cast bound state problems into a
form that can be solved numerically with a reasonable
effort. 

\begin{figure}
\unitlength.4cm
\begin{picture}(14,8.1)(-1,6.8)
\put(1.5,8.5){\line(0,1){1.7}}
\put(1.,11.1){\makebox(0,0){(discrete)}}
\put(1.,12.0){\makebox(0,0){$\perp$ space}}
\put(1.5,12.6){\vector(0,1){1.6}}
\put(2.,8.){\line(3,-1){1.2}}
\put(4.5,6.9){\makebox(0,0){long. space}}
\put(4.5,6.2){\makebox(0,0){(continuous)}}
\put(7.1,6.3){\vector(3,-1){1.8}}
\put(10.4,5.8){\line(3,1){1.8}}
\put(14.5,6.2){\makebox(0,0){(continuous)}}
\put(14.5,6.9){\makebox(0,0){time}}
\put(15.8,7.6){\vector(3,1){1.8}}
\includegraphics{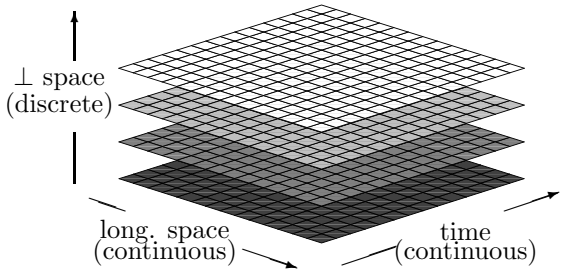}
\end{picture}
\caption{Space time view of a transverse lattice}
\label{fig:perpl}
\end{figure}

The basic idea of the transverse lattice is very 
simple: one keeps two directions (the time and the 
z-direction) continuous but discretizes the 
transverse space coordinates (Fig. \ref{fig:perpl}). 
The metric is Minkowskian.
Two immediate advantages of this construction
are
\begin{itemize}
\item manifest boost and translational invariance
in the longitudinal direction --- thus keeping
parton distributions easily accessible
\item a gauge invariant cutoff for divergences
associated with large transverse momenta
\end{itemize}
Because of these features, the transverse lattice 
seems to be ideally suited for a light-cone
formulation of QCD. 

The gauge degrees of freedom on the transverse 
lattice are described by non-compact $A$-fields in
the continuous longitudinal directions and by
compact link fields $U$ in the transverse directions.
The $A$ fields are defined on the sites of the
lattice and the $U$s on the links. All degrees of
freedom depend on two continuous and two discrete
space-time variables.

For the canonical lattice approximation of
$G_{\mu \nu} G^{\mu \nu}$
on the transverse lattice one needs to distinguish
the following cases:
\bi
\item[1.] Both $\mu$ and $\nu$ are longitudinal.
Here the lattice representation for 
$G_{\mu \nu} G^{\mu \nu}$ is formally
identical to the continuum representation.
\item[2.] Both $\mu$ and $\nu$ are transverse.
Here the lattice representation is just the
plaquette interaction, which is familiar from
Euclidean lattice gauge theory
\item[3.] In the mixed case, i.e. for example
when $\mu$ is longitudinal and $\nu$ is 
transverse, the lattice representation for
$G_{\mu \nu}G^{\mu \nu}$ resembles the kinetic
term of a gauged nonlinear sigma model
\be
G_{\mu \nu}G^{\mu \nu} \longrightarrow
D_\mu U^\dagger D^\mu U.
\ee
\ei
The transverse lattice action was first introduced in
Ref. \cite{bpr} where it was already realized that
a formulation with compact link-fields is not
suitable for light-cone quantization.
\footnote{For a discussion of these technical
difficulties, see for example Refs. 
\cite{paul:zako,mb:korea}.}
This has led to the color dielectric formulation of
the transverse lattice where `macroscopic' or
smeared degrees of freedom are introduced to
represent linearized link degrees of freedom on a 
coarse lattice.
The effective action in the color dielectric
formulation is obtained by making an ansatz which
does not break the unbroken symmetries of the
transverse lattice (e.g. gauge invariance, 
longitudinal boost invariance)
and the coefficients in this ansatz
are then obtained by seeking regions of enhanced
Lorentz symmetry (e.g. where the static $Q\bar{Q}$
potential is `round' and and where light glueballs
have a dispersion relation with the same transverse
speed of light). In Ref. \cite{dalley} such an
ansatz which included terms up to $4^{th}$ order in
the link fields yielded glueball masses that are
consistent with Euclidean results.

\section{Spectrum and structure of light mesons}
We used the effective link-field 
interactions, as determined in previous pure glue 
calculations within the color-dielectric 
formulation of $\perp$ lattice QCD 
\cite{dalley,studs2}.
The fermion action was based on a $\perp$ lattice
generalization of Wilson fermions \cite{mb:hala}. 
In this formulation, there are two hopping terms 
for fermions
(hopping with and without spin flip for the quarks),
which represent also the most general hopping terms
that are possible to the same order in the fields
which are consistent with the residual symmetries 
of the $\perp$ lattice.
\footnote{An alternative formulation, which represents
a generalization of staggered fermions to the $\perp$
lattice, has been described in Ref.  \cite{double}.}

In addition to these hopping terms, one needs to
introduce kinetic energy terms for the quarks as
well as a coupling of the fermions to the 
longitudinal gauge degrees of freedom $A^-$.

Because of gauge invariance, the quarks and $\perp$
link fields couple to the  longitudinal gauge field 
with the same strength. Since gauge coupling for the
$\perp$ link fields have already been determined in 
studies of glueball spectra as well the
static $Q\bar{Q}$ potential, this coupling is no 
longer a free parameter and the only new parameters
are the coefficients of the hopping terms as well as
the (kinetic) quark masses. In the spirit of the
color dielectric formulation, we determined these
parameters by looking for regions in parameter space
with enhanced Lorentz symmetry. 
The criteria that we used to test the 
violation of Lorentz symmetry were
\bi
\item $\kp$ dependence in the dispersion relations
of $\pi$ and $\rho$ mesons: using the relation
$\kp=a\pp$ between the lattice and physical $\perp$
momentum, one can extract the effective $\perp$ 
lattice spacing $a$ for each hadron individually
by performing a Taylor series expansion of its
numerically determined dispersion relation
\be
2p^+p^-_n&=& m_n^2 + c_n^2 \kp^2 + {\cal O}(\kp^4)
\nonumber\\
&=& m_n^2 + a_n^2 c_n^2 \pp^2 + {\cal O}(\pp^4).
\ee
The covariant dispersion relation $2p^+p^-=m^2+\pp^2$
is satisfied if the $\perp$ lattice spacing satisfies
$a_n c_n=1$. The non-perturbative renormalization 
condition that we used in this work was to demand
that the respective lattice spacings for $\pi$ and 
$\rho$ mesons are the same and also agree with 
$\perp$ lattice spacings determined within the pure 
glue sector.
\item mass splitting 
within the $\rho$ meson spin multiplet 
\ei
\begin{figure}
\unitlength1.cm
\begin{picture}(8,3.8)(-7.3,.8)
\includegraphics{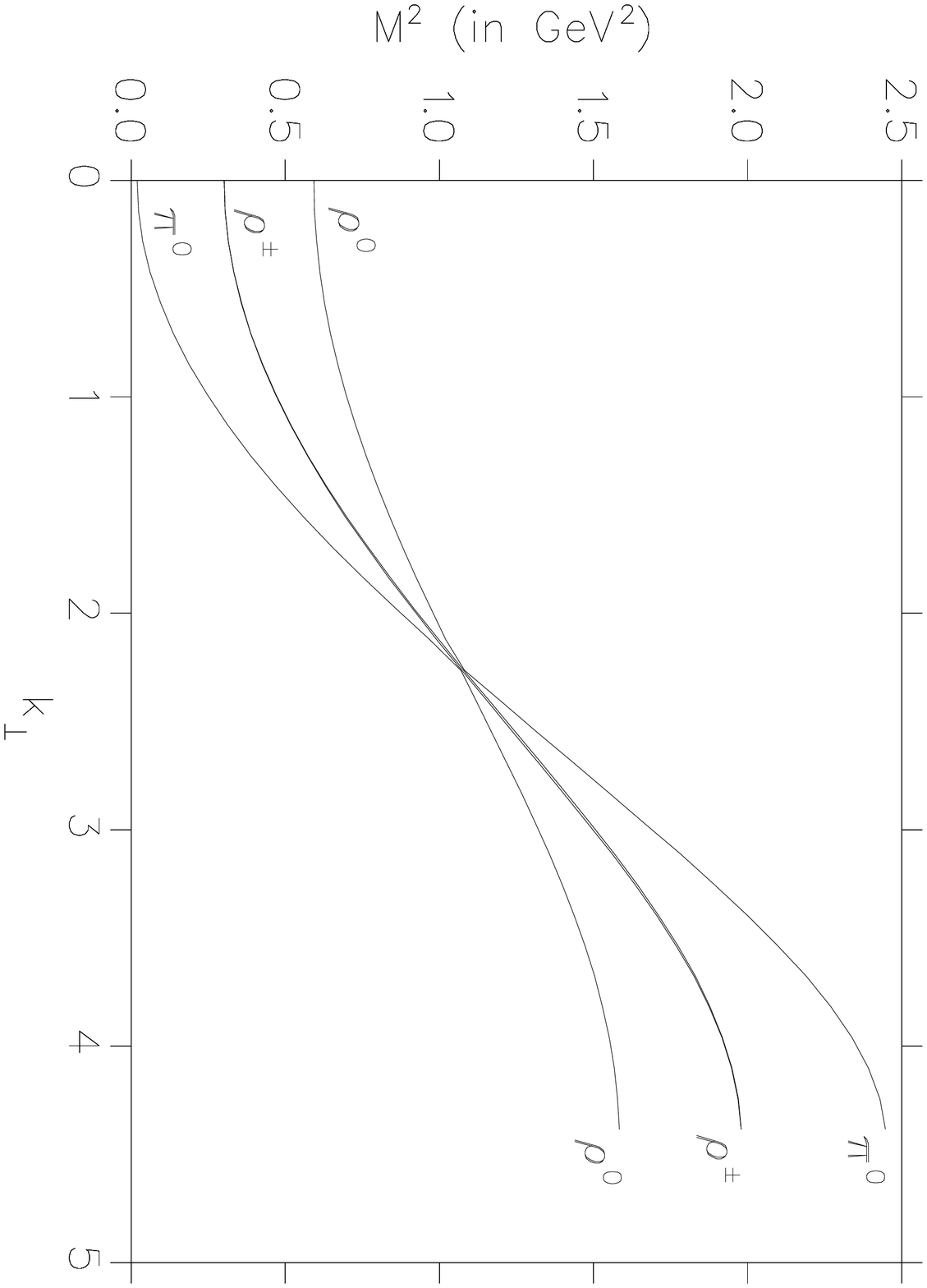}
\end{picture}
\begin{picture}(8,3.5)(-7.3,1.4)
\includegraphics{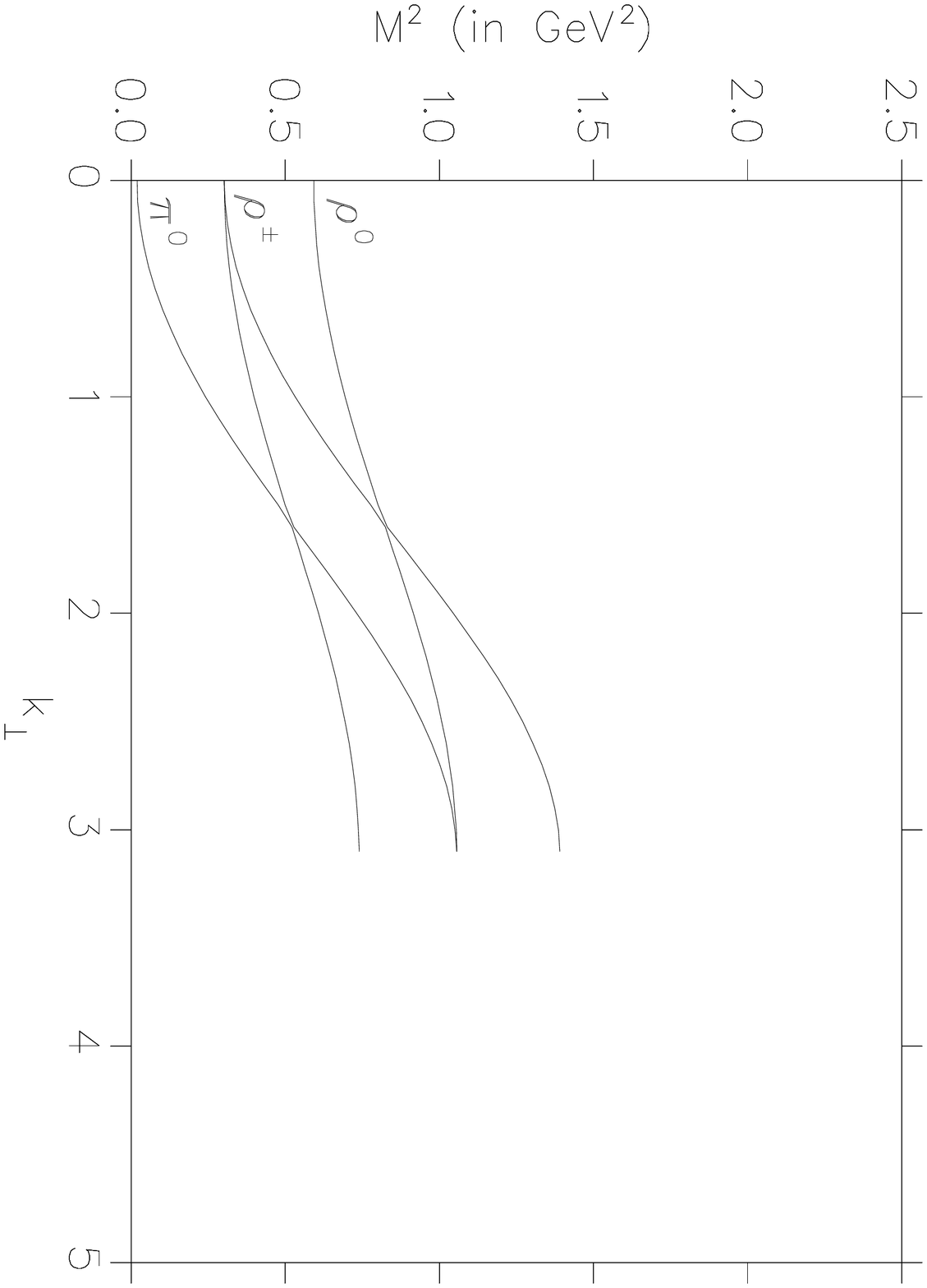}
\end{picture}
\caption{Dependence of the ``transverse mass'' 
$2P^+P^-_n$ of $\pi$ and $\rho$ mesons 
with helicities $0$ and $\pm 1$ on
$\kp$ for momenta along a 
lattice axis and along the diagonal respectively.
}
\label{fig:disp}
\end{figure}
In a first principle calculation, it would be
sufficient to input the $m_\pi$ to set all scales
in the light quark sector. 
%
%
However, since the role of chiral symmetry and
chiral symmetry breaking is still not very well
understood within the LF framework, we needed to
input the chiral symmetry breaking scale as one
phenomenological parameter (on top of the string
tension and $m_\pi$). We did this by using 
$m_\rho$ as an additional input parameter.

In the numerical calculations \cite{sudip1} we 
restricted ourselves
to the $N_C\rightarrow \infty$ limit, which
limited not only the number of possible terms in 
$P^-_{eff}$ 
but also simplified the classification of states.
For the numerical calculations we restricted the
number of link fields to the minimal number which allows $\perp$
propagation of the entire hadron: $q\bar{q}$ on the 
same $\perp$ site as well as $q\bar{q}$ separated
by one link with a link field $U$ connecting
the two.

Typical results for the meson dispersion relations 
are displayed in Fig. \ref{fig:disp}. 
As one can read off from Fig. \ref{fig:disp}, 
it was not possible to restore
full rotational invariance for the $\rho$ multiplet,
but we hope that future calculations including
additional terms in $P^-_{eff}$ as well as higher
Fock components can improve this situation.

In the continuum limit the dispersion relations
should all be parabolas with the same curvature
at the bottom. Near $\kp=0$ this is reasonably
well satisfied, but of course there are larger
violations of Lorentz invariance as the inverse
momentum becomes comparable to the lattice
spacing near the boundaries of the Brillouin zone.
The level crossing is a remnant of species doubling
since without $r$ term species doubling manifests
itself on the transverse lattice at the 
hadronic level by giving rise
to $\rho$ mesons at the boundary that are
degenerate with $\pi$ mesons in the center
of the Brillouin zone (and vice versa).

\begin{figure}
\unitlength1.cm
\begin{picture}(15,4.5)(-8.1,1.5)
\includegraphics{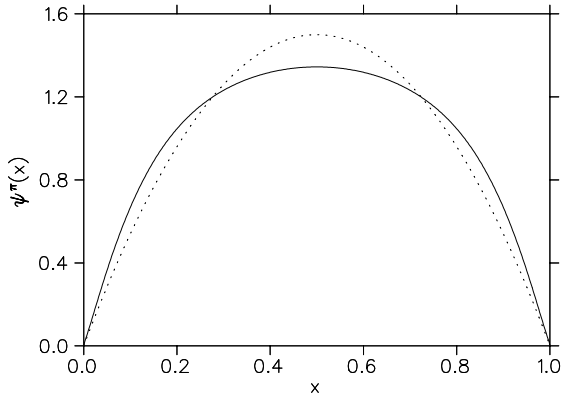}
\end{picture}
\caption{Pion distribution amplitude calculated on
the transverse lattice. For comparison, the 
asymptotic shape is shown as a dashed line.}
\label{fig2}
\end{figure}

Results for the $\pi$ distribution amplitude 
are shown in Fig.
\ref{fig2}. Although $\phi^\pi(x)$ 
resembles the asymptotic distribution
$\phi^\pi_{asy}(x)=6x(1-x)$, our numerically 
determined result is somewhat broader, but clearly
does not exhibit any `double hump' feature.
The $\rho$ meson distribution amplitude looks
similar although it is slightly more peaked, which
reflects the weaker binding of the quarks in the 
$\rho$.

For its normalization, i.e. for $f_\pi$ we found a 
result that is about a factor
2 larger than the experimental value. This 
discrepancy is most likely caused by the Fock space
truncation, since including more higher Fock 
components tends to decrease the probability to 
find a
hadron in its lowest Fock component and therefore
also the normalization of the distribution amplitude.

\section{B mesons on the $\perp$ lattice}
In the limit where the $b$ quark is infinitely heavy,
it acts as a static color source to which the light
quark is bound. The extension of our light meson
calculations to such a heavy-light system is 
straightforward and since the static source does 
not propagate, no new parameters appear in the
Hamiltonian for such a system \cite{sudip2}.

For the decay constant, which also plays an 
important role in mixing phenomenology, we find 
$f_B\approx 240 MeV\pm 20 MeV$, i.e. a value that 
is somewhat larger than those obtained using 
Euclidean lattice gauge theory or QCD sum rules,
but here the discrepancy is much smaller than for
$f_\pi$. This result is consistent with our
observation that the Fock expansion also seems to 
converge much more rapidly for $B$ mesons.

\begin{figure}
\unitlength1.cm
\begin{picture}(15,4.5)(-8.1,1.5)
\includegraphics{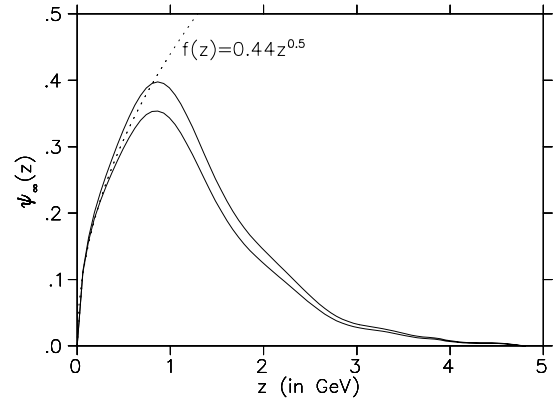}
\end{picture}
\caption{Heavy meson distribution amplitude
in the heavy quark limit. 
The two lines show
the range of numerical uncertainties in the
extrapolation $m_b\rightarrow \infty$.
Dashed line: fit of the endpoint behavior of
the distribution amplitude to 
$\phi_\infty(z)\sim \sqrt{z}$.}
\label{fig:Bdist}
\end{figure}
For phenomenological applications \cite{aa}
it is useful to have numerical estimates for the
moments (normalization: $\int_0^\infty dz
\phi_\infty(z) dz =1$)\footnote{The error bars include only the estimated
error from the numerical extrapolation to the
heavy quark limit and the truncation of the
Hilbert space, but not the systematic errors
from the extrapolation in the Fock space and the
transverse lattice spacing.}
\be
\int_0^\infty dz z \phi_\infty(z)&=&1.51\pm 0.1
\,GeV \nonumber\\
\int_0^\infty 
\frac{dz}{z} \phi_\infty(z)&=&1.22\pm 0.1
\,(GeV)^{-1}.
\ee
The numerical values for the moments indicate a
larger momentum scale compared to other 
calculations. This result is consistent with a rather
large value of the $B$-meson `binding energy'
$\bar{\Lambda}\approx 0.9-1.0\,GeV$ obtained from 
the same transverse lattice eigenstates by
calculating the expectation value of the $p^+$
momentum of all light degrees of freedom in the
$B$ meson. We expect that including more Fock
components will lead to a lowering of 
$\bar{\Lambda}$.

\begin{figure}
\unitlength1.cm
\begin{picture}(15,4.5)(-8.1,1.5)
\includegraphics{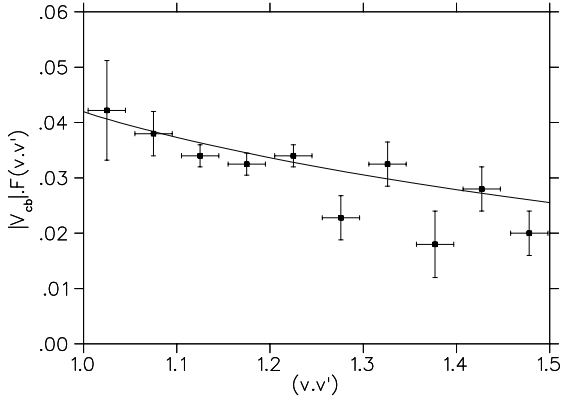}
\end{picture}
\caption{Isgur-Wise form factor (\ref{eq:iwff}) in the heavy quark
limit compared to
experimental data \cite{exp:iwff}.}
\label{fig:iwff}
\end{figure}

An important observable in $B$-physics is the
Isgur-Wise (IW) form factor, because of its use 
in the 
extraction of the CKM matrix element $V_{bc}$ from 
decays like $B\rightarrow \bar{D}^*l \nu$.
We work in the limit $m_c$, $m_b \rightarrow \infty$,
where
$\langle B^\prime |\bar{b}\gamma^\mu b|B\rangle$,
$\langle D^* |\bar{c}\gamma^\mu b|B\rangle$ are
all described by the same universal form factor
\be
\langle B^\prime |\bar{b}\gamma^\mu b|B\rangle
= m_B \left({v^\prime}^\mu+v^\mu\right)
F(v\cdot v^\prime)
\ee
For $m_b, m_c\ll \Lambda_{QCD}$, heavy quark pair
creation is suppressed, i.e. relevant matrix element
is diagonal in Fock space and an overlap representation
exists for $F(v\cdot v^\prime)$
\be
F(v\cdot v^\prime) = F^{(2)} (v\cdot v^\prime) + F^{(3)} (v\cdot v^\prime)
,
\label{eq:iwff}
\ee
where 
\be
F^{(2)} (x)&=& \frac{2}{2-x}\sum_s \int^1_x dz
\psi_s(z)\psi_s^*\left(\frac{z-x}{1-x}\right)
\\
F^{(3)} (x)&=&\frac{2}{2-x}\frac{1}{\sqrt{1-x}}\sum_s
\int_x^1dz\nonumber\\&&\int_0^{1-z}dw\psi_s(z,w)
\psi_s^*\left(\frac{z-x}{1-x},\frac{w}{1-x}\right)
\nonumber
\ee
Here, $\psi_s(z)$ and $\psi_s(z,w)$ are the 
wave functions in the 2 and 3 particle Fock 
component and
$s$ represents the spin/orientation labels.

Numerical results for the shape of the IW form
factor, obtained from our numerically
determined eigenstates on the $\perp$ lattice
are consistent with experimental results.



\begin{thebibliography}{9}
\bibitem{nato} R.J. Perry, lecture notes, 
hep-ph/9710175.
\bibitem{bigguy} S.J. Brodsky, H.-C. Pauli and S.S. 
Pinsky, Phys. Rept. {\bf 301}, 299 (1998);  S.J. 
Brodsky et al., 
Part.\ World\ {\bf 3}, 109 (1993).
\bibitem{osu} K. G. Wilson et al., Phys. Rev. 
{\bf D49}, 6720 (1994).
\bibitem{mb:adv} M. Burkardt, Adv.\ Nucl.\ 
Phys.\ {\bf 23}, 1 (1996).
\bibitem{bpr} W. A. Bardeen and R. B. Pearson, Phys. Rev. {\bf D14}, 547 
(1976); W. A. Bardeen, R. B. Pearson and E. Rabinovici, Phys. Rev. D {\bf 21},
1037 (1980).
\bibitem{paul:zako}P. Griffin, Proc. to `Theory of Hadrons and
Light-Front QCD', Polona Zgorgelisko, August 1994, hep-ph/9410243.
\bibitem{mb:korea} M.~Burkardt, AIP Conf. Proc.
{\bf 494}, 239 (1999); hep-th/9908195.
\bibitem{mbsd} M. Burkardt and S. Dalley,
{\it to appear in} Prog. Part. Nucl. Phys.; 
hep-ph/0112007. 
\bibitem{dalley} S. Dalley and B. van de Sande, Phys.\ Rev.\ Lett.\ {\bf 82}, 1088 (1999),
Phys.\ Rev.\ D\ {\bf 59}, 065008 (1999). 
\bibitem{studs2} B. Klindworth and M. Burkardt, 
in ``Confinement and the Hadron Spectrum 
III'', Jefferson Lab., June 1998, hep-ph/9809283.
\bibitem{mb:hala} M. Burkardt and H. 
El-Khozondar, Phys.\ Rev.\ D\ {\bf 60}, 054504 (1999)
or Phys.\ Rev.\ D\ {\bf 55}, 6514 (1997). 
\bibitem{studs} M. Burkardt and B. Klindworth, 
Phys.\ Rev.\ D\ {\bf 55}, 1001 (1997). 
\bibitem{double} P. Griffin, Phys. Rev. {\bf D47}, 1530 (1993).
\bibitem{sudip1} M. Burkardt and S. Seal, 
{\it to appear in Phys. Rev. D}, hep-ph/0102245.
\bibitem{sudip2} M. Burkardt and S. Seal, 
Phys. Rev. D {\bf 64}, {111501} ({2001}).
\bibitem{aa} M. Beneke and Th. Feldmann,
Nucl. Phys. B {\bf 592}, {3}({2001}).
\bibitem{exp:iwff} CLEO collaboration, 
hep-exp/0007052.
\end{thebibliography}
\end{document}